\documentclass[12pt]{iopart}
\usepackage[utf8]{inputenc}
\usepackage{amssymb,graphicx,booktabs}
\usepackage[colorlinks]{hyperref}
\usepackage[dvipsnames]{xcolor}
\newcommand{\ie}{\emph{i.e.}}
\newcommand{\eg}{\emph{e.g.}}
\newcommand{\modif}[1]{#1}
\begin{document}
\title{Contact network models matching the dynamics of the COVID-19 spreading}
\author{Matúš Medo$^{1,2,3}$}
\address{$^1$ Department of Radiation Oncology, Inselspital, University Hospital of Bern, and University of Bern, 3010 Bern, Switzerland\\
$^2$ Institute of Fundamental and Frontier Sciences, University of Electronic Science and Technology of China, Chengdu 610054, PR China\\
$^3$ Department of Physics, University of Fribourg, 1700 Fribourg, Switzerland}
\ead{matus.medo@unifr.ch}

\begin{abstract}
We study the epidemic spreading on spatial networks where the probability that two nodes are connected decays with their distance as a power law. As the exponent of the distance dependence grows, model networks smoothly transition from the random network limit to the regular lattice limit. We show that despite keeping the average number of contacts constant, the increasing exponent hampers the epidemic spreading by making long-distance connections less frequent. The spreading dynamics is influenced by the distance-dependence exponent as well and changes from exponential growth to power-law growth. The observed power-law growth is compatible with recent analyses of empirical data on the spreading of COVID-19 in numerous countries.
\end{abstract}



\maketitle

\section{Introduction}
Mathematical modeling of epidemic processes has a long tradition~\cite{kermack1927contribution,diekmann2000mathematical,brauer2017mathematical}. After the rise of the network science~\cite{barabasi2016network,newman2018networks}, the study of epidemic spreading on networks has led to a flourishing of literature~\cite{keeling2005networks,barthelemy2011spatial,pastor2015epidemic,porter2016dynamical}. The important insights obtained in this scope include, for example, the impact of network hubs~\cite{pastor2001epidemic} and long-distance connections~\cite{balcan2009multiscale} on the epidemic spreading, and the effective spreading geometry~\cite{brockmann2013hidden}.

By contrast, efforts addressing epidemics on spatial networks with prevailing short connections are comparatively few. In view of the unprecedented movement and activity restrictions that are being introduced in countries across the globe to curb the spreading of the new virus COVID-19~\cite{world2020coronavirus}, this is a research gap that calls to be addressed. Of particular interest are the temporal patterns of the COVID-19 epidemic as there is now extensive empirical evidence that the progress of COVID-19 is power-law in China~\cite{brandenburg2020quadratic,ziff2020fractal,li2020scaling}, Chinese provinces~\cite{maier2020effective,gross2020spatio},
as well as other countries~\cite{gross2020spatio,manchein2020strong}.

Observed power-law growth is very different from the classical exponential growth that follows from the assumption of homogeneous mixing~\cite{kermack1927contribution,diekmann2000mathematical}. To account for this observation, \cite{maier2020effective} introduced a modified SIR model which assumes that the supply of susceptible individuals gradually decreases as a consequence of the implemented containment policies. While the model produces sub-exponential growth and fits the data (prior to February 12, 2020) well, the connection between containment policies which are typically introduced abruptly (such as a curfew issued by a government) and the gradual depletion of susceptible individuals is unclear. On scale-free networks with a diverging second moment of the degree distribution, early power-law growth with the exponent $D-1$ where $D$ is the network diameter (the maximal distance between two nodes in the graph) has been found for the SIR epidemic model~\cite{vazquez2006polynomial}. Even when broad degree distributions are characteristic for social networks~\cite{ahn2007analysis,borgatti2018analyzing}, it is questionable whether a diverging second moment of the degree distribution is a relevant assumption for the contact networks over which COVID-19 spreads in the times of widely-introduced social-distancing rules.

We study here epidemic spreading on model spatial networks introduced in~\cite{medo2006distance} where a single parameter can be used to gradually shift from a random network limit (where links are drawn independently of the spatial node distance and long-range connections are common) to a regular lattice limit (where only the nearest-neighbors are connected). Epidemic spreading in the lattice limit is well understood and characteristic by quadratic growth~\cite{riley2015five,brandenburg2020quadratic} except for the critical regime~\cite{grassberger1983critical}. The spatial network model thus allows us to study the continuous transition between quadratic and exponential growth. A comparison with power-law growth in empirical data, in turn, gives us insights into the contact network over which the epidemic spreads. While contact tracking is the best source of information on the contact network, the contract tracking data are not available now. Investigating the effect of various contact network topologies on the epidemic spreading is hence an option of gaining early insights in the structure of the contact network.

We find that spatial networks where short links are favored facilitate epidemic spreading significantly less than random networks. Together with epidemic characteristics of a considered pathogen, the spatial network structure is thus crucial for the final extent of an outbreak. Furthermore, the epidemic dynamics in these networks features power-law growth with exponents varying in a broad range depending on the network characteristics, in agreement with extensive empirical evidence of power-law growth for the COVID-19 pandemic~\cite{brandenburg2020quadratic,ziff2020fractal,li2020scaling,maier2020effective,gross2020spatio,manchein2020strong}. Our results thus provide an indirect way of probing the structure of actual contact networks over which the virus spreads, as well as suggestions for making these networks less permissible to the virus. Note that we purposely avoid the question of epidemic predictions which has been addressed using phenomenological fits~\cite{jia2020prediction,wu2020generalized,bodova2020emerging} and extensive agent-based models~\cite{chang2020modelling} elsewhere.

\section{Previous work}
The spatial network model that we use here was originally introduced in~\cite{medo2006distance} as a candidate model for social networks. The model is based on connection probability $Q(d)$ which decays with the geometric distance between the nodes. It was shown that when $Q(d)$ decays as a power-law with an exponent between $2.5$ and $3.5$, the resulting networks exhibit the small-world phenomenon: they have simultaneously high clustering and short average shortest paths. In this respect, the spatial decay exponent plays the same role as the rewiring probability in the classical Watts-Strogatz model~\cite{watts1998collective} with small-world networks produced for intermediate values of the network parameter. The degree distribution of networks produced by the original spatial network model is narrow (Poissonian) which markedly differs from degree distributions commonly found in empirical social networks. This shortcoming is addressed in~\cite{medo2008heterogeneous} where nodes are endowed with an additional parameter which quantifies how ``social'' is a given node. A power-law degree distribution then emerges for some distributions of the parameters' values among the nodes whilst maintaining the networks' small-world property.

An epidemic process with spatially distributed sites and space-dependent interaction rates is sometimes referred to as the generalized epidemic process (GEP) in the literature~\cite{grassberger1983critical,janssen2004generalized}. While space-dependent epidemic processes and spatial epidemiology in general~\cite{lawson2016handbook} have a long history, the very basic question of the dynamics of epidemics on spatial networks has been not been fully studied yet. The critical behavior of the GEP was characterized in~\cite{grassberger1983critical}. In~\cite{khaleque2013susceptible}, a SIR model on spatial networks with distance-dependent connectivity has been studied in one dimension. Assuming slowly-decaying spatial interactions, the authors find exponential epidemic growth to best fit their simulation results early on. Epidemic spreading on planar random geometric networks (networks where nodes are randomly distributed in a plane and links exist bellow some node distance) has been studied in~\cite{estrada2016epidemic}, for example, without addressing its temporal dynamics. In~\cite{meyer2014power}, data fits and predictions were compared for epidemic models based on spatial interactions with Gaussian and power-law decay. Using weekly aggregated data covering several years of seasonal variations of meningococcal infections and influenza infections in Germany, the authors found statistical support for a power-law spatial decay of interactions. Besides these works based on individual-level modeling of epidemics, spatially coupled populations have been considered in metapopulation epidemic models~\cite{balcan2009multiscale,hohle2016infectious}.

\section{Preliminaries}

\begin{figure}
\centering
\includegraphics[scale=0.65]{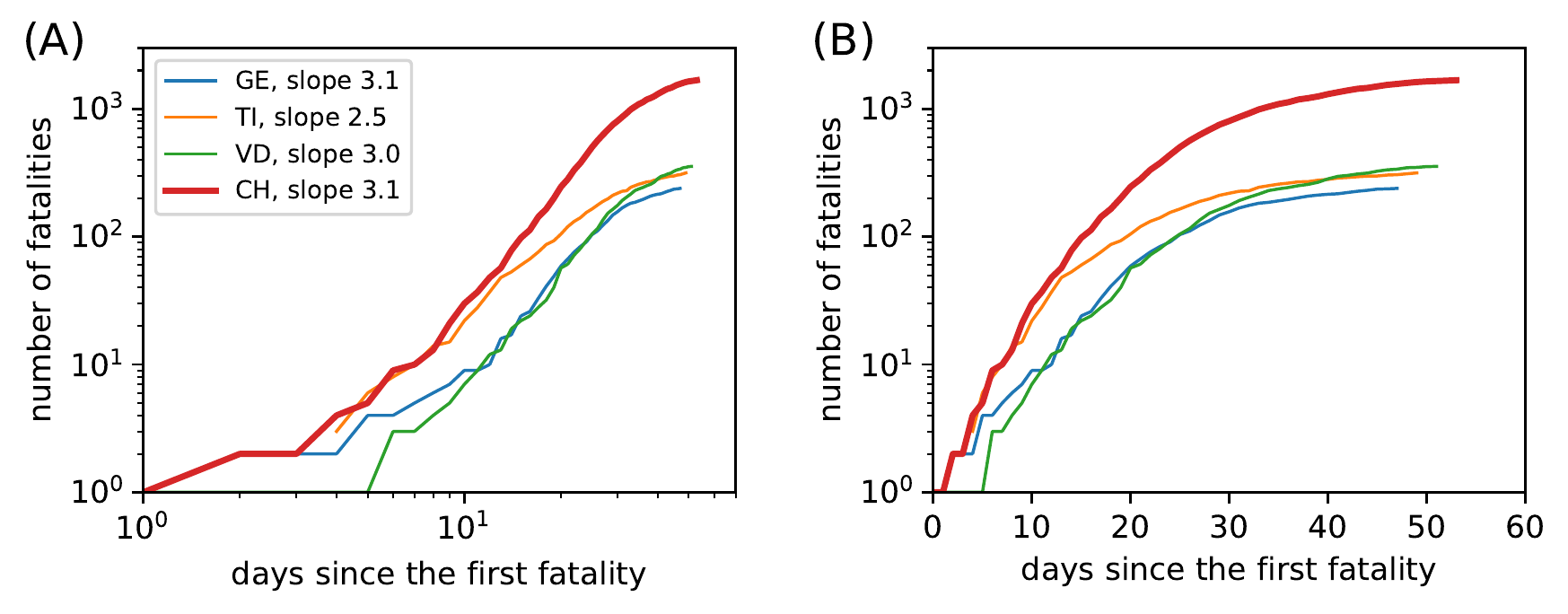}
\caption{The number of COVID-19 fatalities in Switzerland (CH) and three most-affected cantons (Vaud, VD, Ticino, TI, and Geneva, GE) vs. the number of days since the first fatality. See~\cite{brandenburg2020quadratic,ziff2020fractal,li2020scaling, maier2020effective,gross2020spatio,manchein2020strong,jia2020prediction,wu2020generalized,bodova2020emerging} for further empirical evidence of power-law epidemic growth.}
\label{fig:real}
\end{figure}

\subsection{Empirical data}
As we have already mentioned, empirical dynamics of the COVID-19 epidemic has been already studied in a great number of works~\cite{brandenburg2020quadratic,ziff2020fractal,li2020scaling, maier2020effective,gross2020spatio,manchein2020strong,jia2020prediction,wu2020generalized,bodova2020emerging}. To provide a direct motivation for our study of power-law patterns in epidemic dynamics, we offer here a simple vizualization of empirical data on COVID-19 fatalities in Switzerland. We choose the fatalities count as the other candidate characteristic, the number of confirmed cases, is affected by issues such as asymptomatic cases, the extent of testing, and the choice of the tested individuals. We utilize the data used by the website \url{http://www.corona-data.ch/}; the data have been collected and published under the OpenZH program of the Swiss government (see \url{https://open.zh.ch}). The data covers the period from the first confirmed COVID-19 case in Switzerland (February 25, 2020) until April 27, 2020. As can be seen in Figure~\ref{fig:real}, the empirical curves are straight over a substantial part of the reported period in the log-log plot (panel A), suggesting a power-law growth pattern, instead of being straight in the log-linear plot (panel B) which would suggest an exponential growth pattern. As a direct consequence of strict country-wide social distancing rules (\eg, all meetings over 5 people were banned on March 20), the growth slows down substantially after day 30, approximately.

\subsection{Spatial network model}
\label{sec:model}
In a spatial complex network, the probability that two nodes are connected is given by the distance between them~\cite{barthelemy2011spatial}. We use here the model introduced in~\cite{medo2006distance} where $N$ nodes form a two-dimensional square lattice with the elementary square of size one and periodic boundary conditions; we assume for simplicity that $N=S^2$ where $S$ is a natural number. In this model, the probability that two nodes are connected depends only on their geometric distance $d$, hence the original label ``network with distance-dependent connectivity'', as
\begin{equation}
\label{P(d)}
P(d) = \frac1{1 + \lambda d^{\delta}}
\end{equation}
where $\lambda$ and $\delta$ are model parameters. The exponent $\delta$ determines how fast $P(d)$ decays with distance and $\lambda$ can be used to achieve a desired mean degree, $z$, in the resulting network. When $P(d)$ is small, it changes slowly with $d$, and $\delta > 2$, mean node degree can be approximated as
\begin{equation}
\label{z_approx}
z\approx\int_0^{\infty} 2\pi rP(r)\,\mathrm{d} r = \frac{2\pi^2\lambda^{-2/\delta}}{\delta\sin(2\pi/\delta)}
\end{equation}
in the limit of $N\to\infty$~\cite{medo2008heterogeneous}. For the sake of generality, we always determine $\lambda$ necessary to achieve a desired mean degree numerically by evaluating the mean degree in generated networks, even for parameter settings where $\lambda$ determined from \eref{z_approx} would be of sufficient precision.

By contrast to the simpler form $P(d)\sim d^{-\delta}$ used in~\cite{daqing2011dimension}, the absolute term in the denominator conveniently avoids the divergence of $P(d)$ when $d\to0$. In~\cite{medo2006distance}, the authors show that \eref{P(d)} with $2.5\lesssim\delta\lesssim 3.5$ produces small-world networks where the average shortest paths are ``short'' and the clustering coefficient values are high. The model thus formalizes what has been anticipated by previous analyses of the COVID-19 data: ``individuals have many local neighbors and occasional long-range connections''~\cite{ziff2020fractal}. See~\cite{meyer2014power} for further evidence for the use of power-law decay in epidemics modeling.

\begin{figure}
\centering
\includegraphics[scale=0.65]{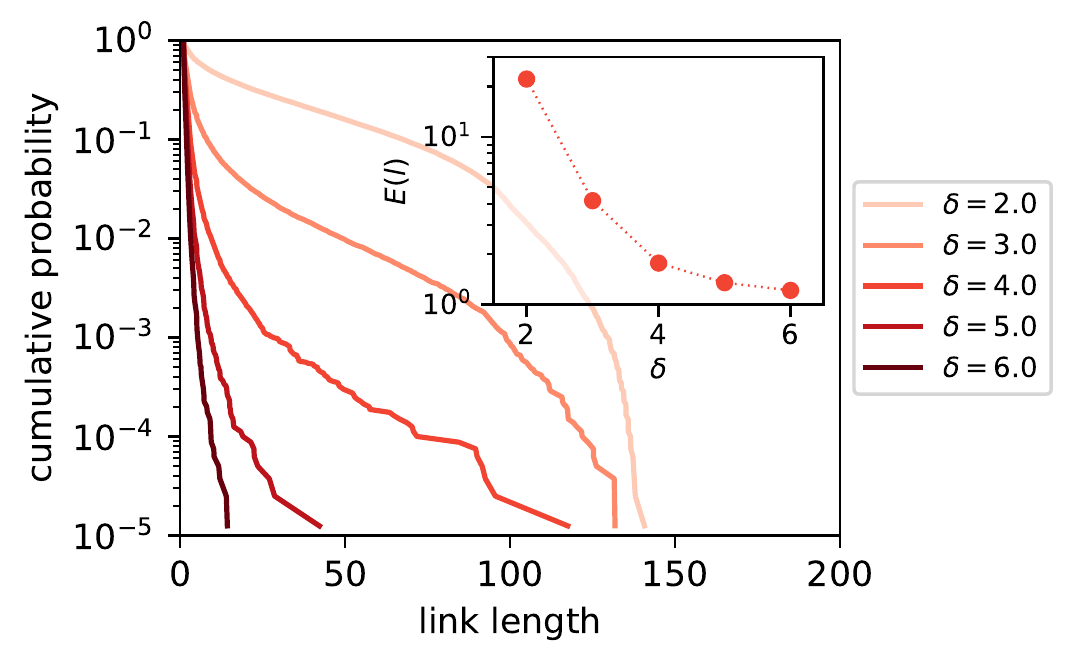}
\caption{The distribution of link lengths in model spatial networks with $N=40,000$. For each $\delta$, $\lambda$ is chosen to achieve the mean degree $z=4$. The inset shows the mean link length, $E(l)$, as a function of the exponent $\delta$. For $N=40,000$, the longest possible link is $\sqrt{N}\sqrt{2}/2\approx141$ and the mean length of random links is $77$.}
\label{fig:DDC}
\end{figure}

Figure~\ref{fig:DDC} shows that as $\delta$ grows, links in model networks become shorter. When $\delta\gg1$, short links have much higher probability and the network is nearly regular as the nodes are mostly connected with their $z$ closest neighbors. The opposite limit, $\delta\to 0$, is also instructive: node distance then becomes irrelevant and the connection probability is the same, $1/(1 + \lambda)$, for all nodes. We thus recover the classical random network~\cite{bollobas2001random}.

The original spatial model produces a narrow degree distribution which is typically not a good fit for real social networks~\cite{ahn2007analysis,borgatti2018analyzing}. This limitation can be overcome by assigning individual $\lambda$ values to all nodes and making \eref{P(d)} dependent on the $\lambda$ values of the considered pair of nodes~\cite{medo2008heterogeneous}. We use
\begin{equation}
\label{P(d)_hetero}
P(d_{ij}) = \frac1{1 + \min(\lambda_i, \lambda_j) d^{\delta}}
\end{equation}
as the probability that nodes $i$ and $j$ with distance $d_{ij}$ and respective values $\lambda_i$ and $\lambda_j$ are connected. Note that a smaller $\lambda$ value in \eref{P(d)} leads to a higher connection probability. By using the minimum $\lambda$ value in \eref{P(d)_hetero}, we thus make it possible that a node with a small $\lambda$ value connects with many nodes in the network. Albeit this choice slightly differs from the one used in~\cite{medo2008heterogeneous}, the tail behavior of the degree distribution remains unchanged. In particular, $\lambda$ distributed among the nodes as $\varrho(\lambda)\sim\lambda^x$ in $\lambda\in(0,\lambda_{\max}]$ has been shown in~\cite{medo2008heterogeneous} to lead to a power-law tail of the degree distribution with the exponent $\gamma=\delta(x+1)/2$. The upper bound of the $\lambda$ distribution, $\lambda_{\max}$, directly determines the mean degree. We use this approach to generate spatial networks with the degree distribution exponents $3.5$ and $2.5$, respectively. We use the basic network model characterized by a single $\lambda$ value and \eref{P(d)} unless stated otherwise.

\subsection{Epidemic model}
We use a SEIR epidemic model~\cite{diekmann2000mathematical,hethcote2000mathematics} where each node can be in one of four possible states: susceptible (S), exposed (E), infected (I), and recovered (R). All nodes are initially susceptible except for one node which is exposed. The exposed nodes represent the individuals who have already contracted the disease but have not developed the symptoms yet. In the case of COVID-19, these individuals have been shown to significantly contribute to spreading the disease~\cite{li2020substantial}.

The epidemic takes place on a fixed contact network and the simulation runs in time steps with one step representing one day. If susceptible node $i$ is connected with an exposed node, the probability that node $i$ becomes exposed is $\beta_1$. If susceptible node $i$ is connected with an infected node, the probability that node $i$ becomes exposed is $\beta_2$. In the simulations, we go over all nodes in a random order in each turn and if a node is exposed or infected, we make its susceptible neighbors exposed with probabilities $\beta_1$ or $\beta_2$, respectively. While the infection process is probabilistic, we assume for simplicity that the disease progression is deterministic: If a node becomes exposed at time step $t$, it automatically becomes infected (develops disease symptoms) at the end of step $t + T_1$ and recovered at the end of step $t + T_1 + T_2$. Here $T_1$ is the incubation period and $T_2$ is the recovery time. Recovered nodes cannot contract the disease again.

We use $T_1=5$, $T_2=14$, and $\beta_2/\beta_1 = 0.2$ which reflects the role of asymptomatic agents in the spreading process~\cite{li2020substantial}. When $\beta_1$ is sufficiently small, the probability that a node infects a given neighbor is $\beta_1T_1$ while in the $E$-state and $\beta_2T_2$ while in the $I$-state. For the specified parameter values, the probability of infecting a neighbor while in the $E$-state is $5\times 5/14\approx 1.8$ fold with respect to the $I$-state which is in the range suggested by the empirical literature (see \cite{li2020early,bar2020sars} for the latest epidemiology information on COVID-19). While the choice of the model and its parameters are motivated by the ongoing COVID-19 epidemic, the presented findings are robust with respect to the choice of the epidemic model and the parameter values. Note that we study specifically the super-critical regime where the disease spreads through the population and eventually reaches a significant fraction of the whole population. This is different from, for example, \cite{grassberger1983critical} where the critical behavior of epidemic spreading was addressed.

We simulate 300 time steps of the epidemic spreading. Median over 10,000 independent realizations of the epidemic process is used to find a representative epidemic outcome at any time point. We measure the epidemic spreading using the number of affected nodes which includes exposed, infected and recovered nodes, so it is naturally a cumulative metric which quantifies how many nodes have contracted the disease during the simulation time.

\begin{figure}
\centering
\includegraphics[scale=0.65]{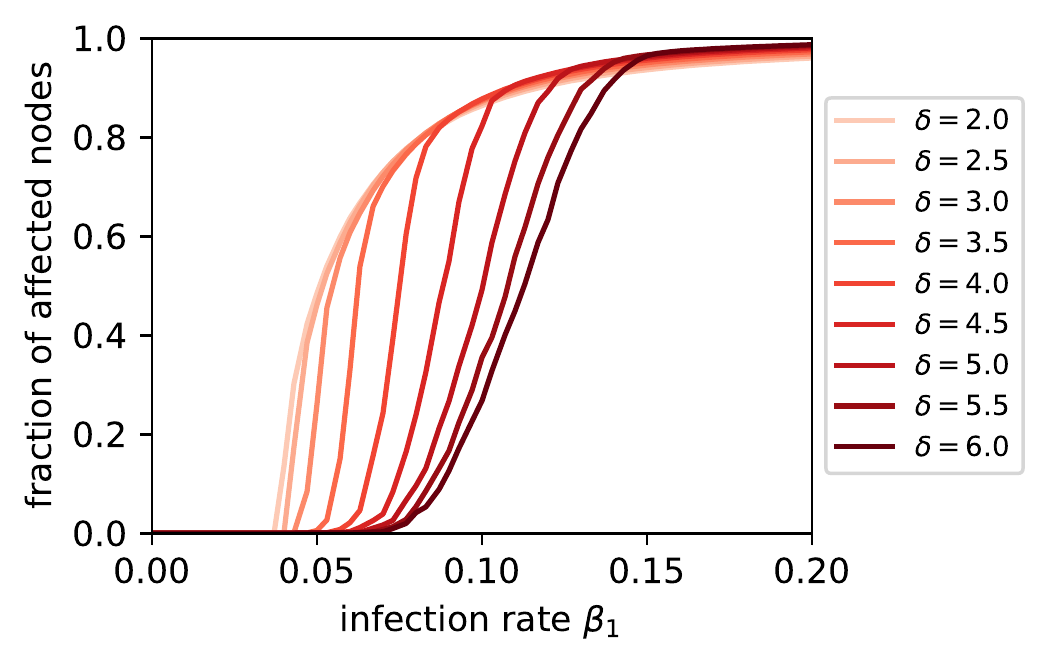}
\caption{The fraction of affected nodes (the sum of exposed, infected, and recovered nodes) at $t=300$ for $N=40,000$ and $z=4$ as a function of the infection rate.}
\label{fig:final_vals}
\end{figure}

\section{Results}
We now move to results of numerical simulations of epidemic spreading on model spatial networks. Before addressing the dynamics of the epidemic process, Figure~\ref{fig:final_vals} shows that the spatial distribution of links, controlled in the model by the exponent $\delta$, has strong impact on the epidemic spreading. In particular, the fraction of infected nodes at a given time decreases as $\delta$ grows. This is a direct consequence of the spatially-constrained epidemic spreading when $\delta$ is large: Many neighbors of an infected node are then already infected, so the node's effective ability to spread the disease further is reduced. By contrast, neighbors of an infected node in a random network are unlikely to be already infected (unless a significant fraction of all nodes have been infected) which facilitates further spreading. In summary, the epidemic spreading can be effectively suppressed by both lowering the number of contacts (in the case of COVID-19, in particular close contacts that we spend prolonged time together indoors~\cite{qian2020indoor}) as well as by avoiding long-distance contacts. A comparatively higher number of short-distance social contacts can lead to the same, or even lower, final fraction of infected population than a lower number of widely distributed social contacts.

\begin{figure*}
\centering
\includegraphics[scale=0.65]{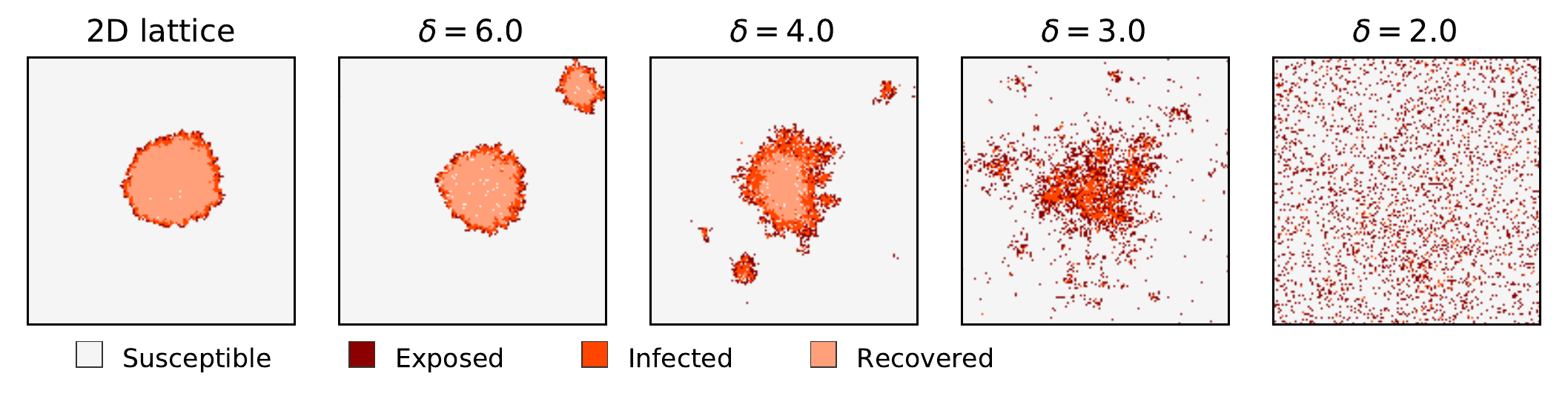}
\caption{Epidemic spreading on the square lattice (first panel) and model spatial networks with various values of the exponent $\delta$ (following panels). We use $N=160,000$ and $z=4$. The snapshots are taken when 10\% of all nodes have been affected by the epidemic. We use $\beta_1=0.16$ for which half of all nodes become affected at the end of simulation for $\delta=6$.}
\label{fig:interim}
\end{figure*}

The effect of $\delta$ can be readily understood by inspecting the spreading patterns for various values of $\delta$ shown in Figure~\ref{fig:interim}. When $\delta$ is high ($\delta=6$), the epidemic spreads almost as on a regular lattice (see the first panel) and develops a clearly-defined spreading front. Nodes on the front, that can spread the infection further, can only spread it in one direction (outwards) which limits their effective reproduction number (number of nodes that they infect on average). As $\delta$ decreases, the spreading front first becomes ``diffuse'' ($\delta=4$) until it ``dissolves'' entirely ($\delta=2$). The epidemic then spreads more effectively as it is more probable that a neighbor of an infected nodes is susceptible to the infection. \modif{The qualitative change from a spatially localized spreading (when $\delta$ is high) to a delocalized spreading (the last panel in Figure~\ref{fig:interim}) occurs when the average link length, $E(l)$, diverges (\ie, for $\delta\leq 3$). As the infection spreads over network links, the geometric speed of infection spreading then diverges too.}

We now return to the main question: the effect of the spatial network structure on the dynamics of epidemic spreading. This dynamics has two well-known special cases. On a random contact network, the assumption of homogeneous mixing of susceptible and infected nodes is valid and one recovers the classical exponential epidemic growth~\cite{diekmann2000mathematical}. On a regular two-dimensional lattice with nearest-neighbor connections, the epidemic spreading instead develops an epidemic front that propagates at a constant velocity. The front's constant velocity then directly implies that the radius of the affected area grows linearly with time and, consequently, the number of infected individuals grows quadratically with time~\cite{riley2015five}.

\begin{figure*}
\centering
\includegraphics[scale=0.65]{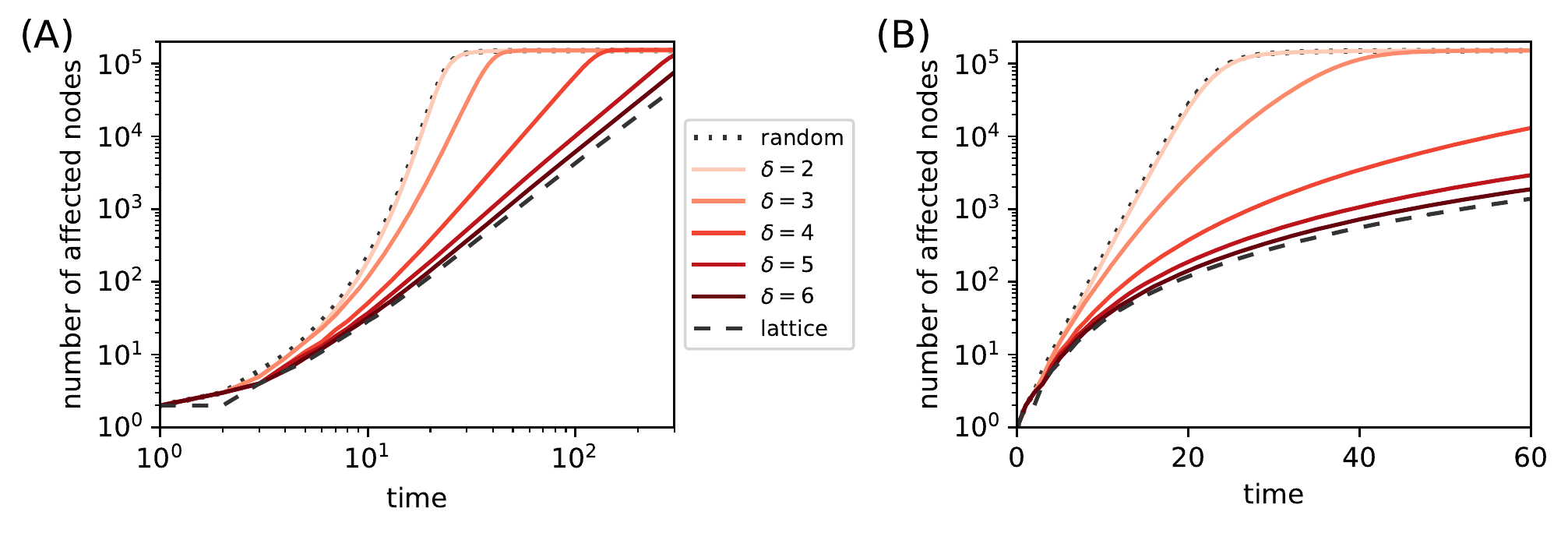}
\caption{Time evolution of the number of affected nodes for various contact networks. While the solid lines show results for spatial networks generated with different exponent values in \eref{P(d)}, the dotted and dashed lines show results for a random network and a regular lattice, respectively. Straight lines in the log-log panel (A) are indicative of power-law growth. Straight lines in the log-linear panel (B) are indicative of exponential growth. As in Figure~\ref{fig:interim}, we use $N=160,000$, $z=4$, and $\beta_1=0.160$. We display here a median of 10,000 epidemic realizations on a single network.}
\label{fig:evolution}
\end{figure*}

Spatial networks with distance-dependent connectivity give us the possibility to smoothly transition between the two extremes (regular lattice and classical random network). We can thus directly observe how quadratic growth on a lattice changes in exponential growth on a random network. Figure~\ref{fig:interim} provides the first indication with the epidemic front becoming more diffuse and the speed of its propagation grows as $\delta$ decreases. As this happens, the argument leading to quadratic growth ceases to be valid and we expect the growth to be faster than quadratic. This is confirmed by Figure~\ref{fig:evolution} where we show the epidemic dynamics for various contact networks. Growth that is of a power-law kind (it follows a straight line over a substantial part of the log-log plane in panel A) for the 2D lattice as well as for the spatial networks with $\delta\gtrsim 4$ becomes clearly exponential for the random network and for the spatial network with $\delta = 2$ (see the straight lines in the log-linear panel B). Note that the results obtained on random networks and the regular square lattice are very close to the displayed results for $\delta=2$ and $\delta=6$, respectively.

\begin{table}
\caption{\label{tab:fits} Power-law exponents, $\hat b$, estimated by fitting straight lines to the data in Figure~\ref{fig:evolution}A. To focus on the straight part of the dependencies, we fit only the values below $N/2$ infected cases and above $100$--$800$ infected cases. The reported ranges of the fitted exponents are obtained by varying the fitting lower bounds. The exponent estimates for $\delta=3$ vary in comparatively broader ranges, indicating that power-law growth is not a good fit in this case (the same is true for $\delta=2$, results not shown). The first two rows show results for the basic spatial network model with different mean degree values, $z$. The next row shows results for the spatial network model with a power-law degree distribution with the exponent $\gamma=3.5$.}
\centering
\begin{tabular}{rrrrr}
\toprule
$\delta$ & 3 & 4 & 5 & 6\\
\midrule
           $\hat b(z = 4)$ & 5.25--5.60 & 3.23--3.26 & 2.42--2.44 & 2.29--2.31\\
          $\hat b(z = 10)$ & 5.52--5.86 & 3.53--3.56 & 2.51--2.56 & 2.33--2.38\\
$\hat b(z = 4,\gamma=3.5)$ & 5.67--5.87 & 3.65--3.67 & 2.67--2.69 & 2.44--2.45\\
\bottomrule
\end{tabular}
\end{table}

\begin{figure*}
\centering
\includegraphics[scale=0.65]{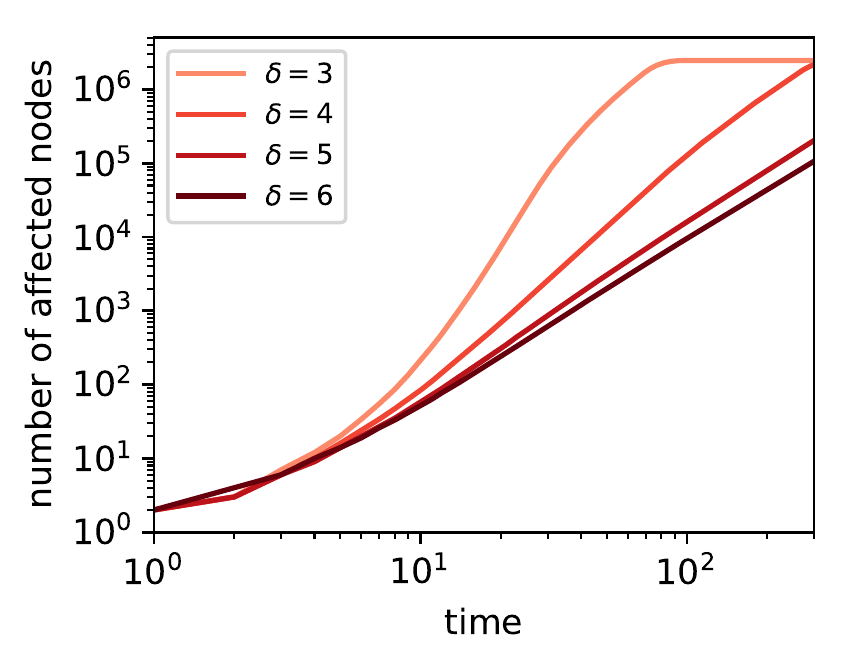}
\caption{Time evolution of the number of affected nodes for large spatial networks (log-log scale). We use $N=2,560,000$, $z=4$, and $\beta_1=0.160$. We display here a median of 1,000 epidemic realizations on a single network.}
\label{fig:larger}
\end{figure*}

We find that quadratic epidemic growth on 2D lattices generalizes to power-law growth on the model spatial networks with sufficiently high exponent $\delta$. Table~\ref{tab:fits} shows that for the data shown in Figure~\ref{fig:evolution}A, stable estimates of the growth exponent (indicating good power-law fits of the data) can be obtained for $\delta$ greater or equal to 4. The resulting power-law growth exponents supported by simulations with $N=160,000$ nodes lie in the range $2$--$3.25$ (for $\delta=3$, power law ceases to fit well). This range of exponents agrees well with the exponents found in empirifcal data~\cite{ziff2020fractal,li2020scaling,maier2020effective,gross2020spatio,manchein2020strong}. A comparison with the results obtained for $N=40,000$ (results not shown) suggests that as $N$ grows, the range of $\delta$ where a power-law fits the data well expands. Notably, the power-law exponent is more than two even for the square lattice (for the same fitting procedure as in Table~\ref{tab:fits}, we obtain 2.17--2.19). This is due to the probabilistic nature of the spreading that results in non-vanishing randomness of the spreading front that can be well seen in the first panel of Figure~\ref{fig:interim}. Simulations at larger $N=2,560,000$ (Figure~\ref{fig:larger}) show analogous behavior, confirming that the observed growth patterns are not significantly influenced by finite-size effects. The fitted slopes 5.4, 3.2, 2.4 and 2.2  for $\delta=3,4,5,6$, respectively, agree with the slopes obtained for $N=160,000$. Since the original network construction scales poorly as it requires checking all possible pairs of nodes which results in complexity $O(N^2)$, we accelerate it in this case by checking only pairs of nodes whose distance is less than $200$ (all other simulation aspects and parameters remain the same). In this way, only a minor fraction of long links is omitted for $\delta\geq3$ shown in Figure~\ref{fig:larger}. We finally note that the considered epidemic spreading model assumes no fundamental difference between infections by exposed nodes and infected nodes (except for the infection rate, of course), so the ratio $T_1/T_2$ does not influence the results despite affecting the relative proportion of exposed and infected nodes.

In~\cite{manchein2020strong,singer2020short}, the authors observe power-law growth in various countries and infer that the underlying networks must be scale-free and small-world (the small-world property is hypothesized also in~\cite{li2020scaling,ziff2020fractal}). We now see that this is not necessarily the case: The spatial networks considered here can produce power-law growth when $\delta\gtrsim4$ yet they have narrow degree distributions~\cite{medo2008heterogeneous} (\ie, they are not scale-free) and \cite{medo2006distance} shows that high exponents in the distance dependence do not generate sufficiently many long-range links for the small-world phenomenon to emerge. We find instead that power-law growth with an exponent above two suggests that the contact network is spatially embedded and consists predominantly of short links.

\begin{figure*}
\centering
\includegraphics[scale=0.65]{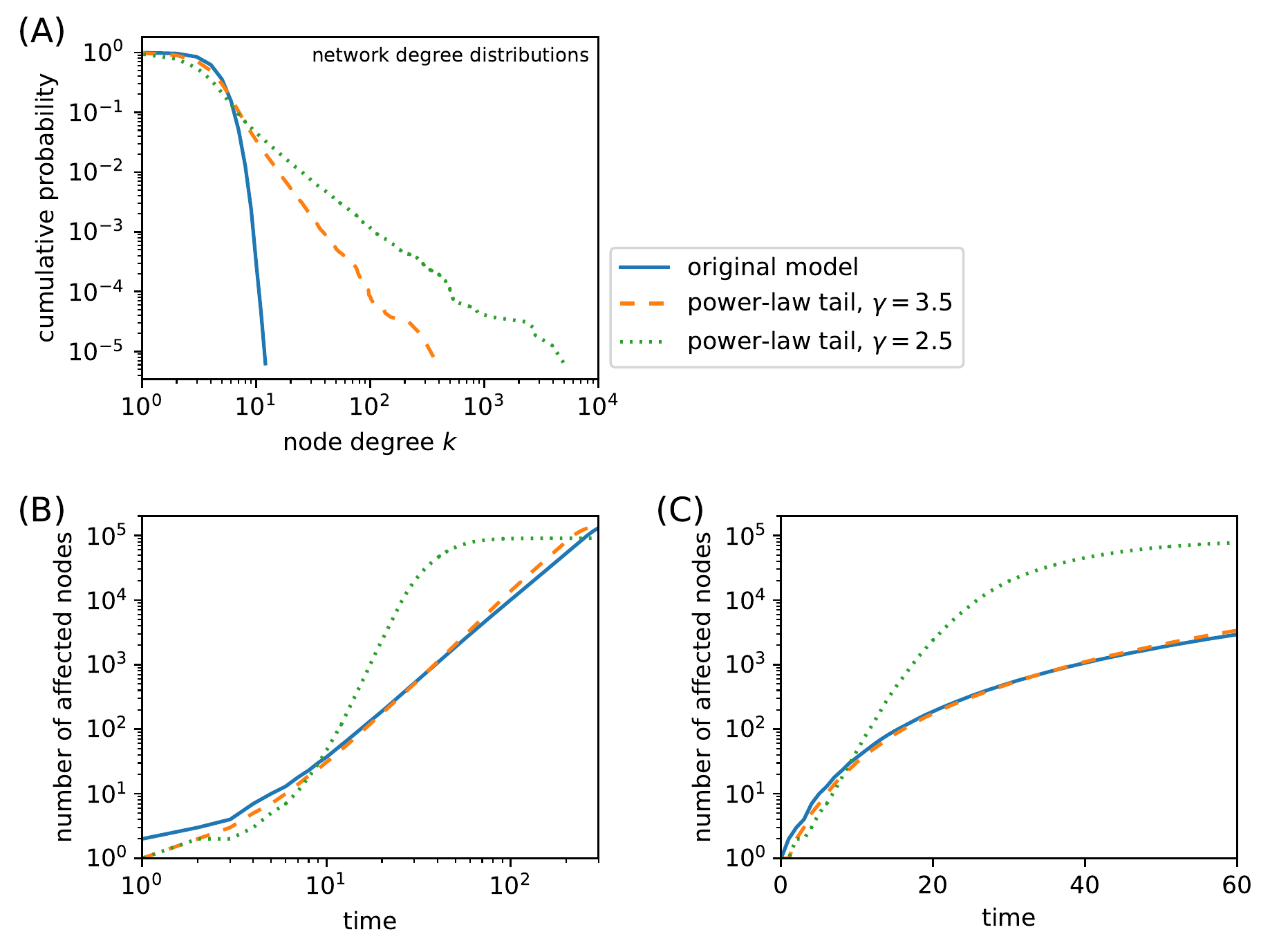}
\caption{Results for spatial networks with various degree distributions: original model with a narrow degree distribution and power-law distributions with target exponents $\gamma$ 3.5 and 2.5, respectively. (A) Networks degree distributions. Maximum likelihood fits~\cite{clauset2009power} of the two broad distributions yield the exponents 3.50 and 2.51 for degree or more, respectively, thus validating our network construction and the analytical result for exponent $\gamma$ provided in Section~\ref{sec:model}. (B, C) The epidemic dynamics in the log-log and log-linear layout, respectively. We use $N=160,000$ and $z=4$; representative results obtained with $\delta=5$ are shown. The transmission rates are $\beta_1=0.160$ (original model), $\beta_1=0.116$ ($\gamma=3.5$), and $\beta_1=0.063$ ($\gamma=2.5$) for which, as we already used in Figure~\ref{fig:interim}, half of all nodes become affected at the end of simulation when $\delta=6$.}
\label{fig:summary}
\end{figure*}

We finally explore how variations of the underlying network change the estimated power-law growth exponents, $\hat b$ (see Table~\ref{tab:fits} for the detailed results). When the mean degree increases, $\hat b$ increases and this increase is more pronounced (up to 10\%) for small $\delta$ values. This is actually expected as higher $z$ means that nodes on the epidemic front have more opportunities to have long-distance connections which, in turn, accelerates the epidemic front's movement forward. An increase of $\hat b$ can be also observed when \eref{P(d)_hetero} is used to produce spatial networks with a power-law degree distribution. When the degree distribution exponent is $\gamma=3.5$, $\hat b$ increases by up to 13\% with respect to the original model with $z=4$. Interestingly, power-law epidemic growth then appears to fit the data even better (\eg, the exponent ranges in Table~\ref{tab:fits} are narrower) than for the original networks with narrow degree distributions. While the networks' fractal dimensions~\cite{daqing2011dimension} are similar to the power-law exponents obtained by fitting the epidemic growth, substantial differences (\eg, fractal dimension between 3.9 and 4.0 for $N=160,000$, $z=10$ and $\delta=4$ as opposed to the epidemics growth exponent, $\hat b$, around $3.55$ for the same parameters) can be observed in some cases. These differences stem from the fact that while the evaluation of a network's fractal dimension is deterministic, it uses only on the network's structure, the epidemic dynamics is inherently stochastic, driven by both the network structure as well as details of the studied epidemic process. Differences between the fractal dimension and the epidemic growth exponents are expected to further magnify when link weights, affecting the epidemic spreading but irrelevant to the computation of the fractal dimension, are introduced.

The situation becomes very different when the degree distribution exponent $\gamma$ is 2.5: the power-law parts of the epidemic growth then have higher exponents ($4.2$ for $\delta=6$ and $5.9$ for $\delta=5$). However, the range over which power-law growth can be observed shrinks substantially and exponential growth becomes a better description for the growth pattern. This is illustrated in Figure~\ref{fig:summary} which shows a direct comparison of the growth dynamics on original spatial networks and spatial networks with power-law degree distributions. The change of behavior for $\gamma=2.5$ is a direct consequence of a diverging second moment of the degree distribution, $\langle k^2\rangle$, for a power-law exponent below $3$; $\langle k^2\rangle$ has been shown crucial for the dynamics of epidemic spreading on networks with heterogeneous degree distributions~\cite{pastor2001epidemic}. Results remain qualitatively the same when a different infection rate $\beta_1$ is used.

\section{Discussion}
In this work, we studied the interplay between the functional form of the connection probability in spatial networks and the dynamics of epidemic spreading on them. There are two main findings. Firstly, short (localized) links hinder the spreading. A better outcome (fewer infections and slower epidemics growth) can be thus achieved by limiting distant contacts even if the mean number of contacts remains fixed. Secondly, when short links prevail in the network, the number of infected individuals grows as a power law instead of the canonical exponential epidemic growth. In the framework of the chosen connection probability decaying with node distance as a power law, frequent short links are achieved when the exponent of the connection probability decay is sufficiently high. In our simulations, in particular, power-law growth of the number of infections emerges when this decay exponent is greater than, approximately, four. This observation is particularly relevant as the latest data show that the COVID-19 spreading in most countries indeed follows a power-law pattern instead of exponential growth. \modif{Our results point to short (spatially) social contacts as a candidate mechanism behind the observed power-law growth of COVID-19.}

We stress that the studied mechanism leading to power-law growth of the number of infections is very different from the setting studied in~\cite{vazquez2006polynomial} where a diverging second moment of the degree distribution is the main reason for sub-exponential growth. It remains open whether the degree distribution of the effective contact network over which COVID-19 spreads has a diverging second moment, in particular as a major epidemic outbreak necessarily affects the behavior of individuals and incites significant government interventions and society-wide restrictions. By contrast, the original spatial model that we consider here produces narrow (Poissonian) degree distributions, yet we find power-law epidemic growth as a consequence of spatially-constrained linking patterns which, in turn, determine the spreading dynamics. \modif{Our results thus show that neither the scale-free property (hypothesized in~\cite{manchein2020strong,singer2020short}), nor the small-world property (hypothesized in~\cite{li2020scaling,ziff2020fractal}) are necessary for power-law epidemics growth to emerge.}

Note that the result obtained in \cite{vazquez2006polynomial} includes power-law growth combined with exponential decay. While this exponential decay term can be conveniently used to fit the slowing epidemic data~\cite{bodova2020emerging}, the source of this decay is the shrinking size of the susceptible population which, fortunately, is not a significant effect for COVID-19 which has yet affected a minor fraction of the world's population. In the scope of fitting the COVID-19 data, the exponential decay term has to be thus viewed as phenomenological.

To better understand the impact of various features of the contact network on the resulting epidemic spreading, several modifications and generalizations of the network model can be studied in the future: (1) replacing individual nodes with households where each node has local connections to all other household members and distance-dependent connectivity is used to model the connections to individuals in other households, (2) introducing link weights that represent the intensity of the social contact and naturally play a role in the disease transmission, (3) \modif{consider interventions and social distancing that vary in time}. Finding an analytical relation between the epidemic growth exponent and the connectivity decay exponent $\delta$ in the basic model, and formulating effective differential equations that describe the epidemic dynamics on spatial networks are other important directions.

\section*{References}
\bibliographystyle{iopart-num}
\bibliography{refs_epidemics}

\providecommand{\newblock}{}
\begin{thebibliography}{10}
\expandafter\ifx\csname url\endcsname\relax
  \def\url#1{{\tt #1}}\fi
\expandafter\ifx\csname urlprefix\endcsname\relax\def\urlprefix{URL }\fi
\providecommand{\eprint}[2][]{\url{#2}}

\bibitem{kermack1927contribution}
Kermack W~O and McKendrick A~G 1927 {\em Proceedings of the Royal Society of
  London, Series A\/} {\bf 115} 700--721

\bibitem{diekmann2000mathematical}
Diekmann O and Heesterbeek J~A~P 2000 {\em Mathematical epidemiology of
  infectious diseases: {M}odel building, analysis and interpretation\/} vol~5
  (John Wiley \& Sons)

\bibitem{brauer2017mathematical}
Brauer F 2017 {\em Infectious Disease Modelling\/} {\bf 2} 113--127

\bibitem{barabasi2016network}
Barab{\'a}si A~L {\em et~al.\/} 2016 {\em Network science\/} (Cambridge
  University Press)

\bibitem{newman2018networks}
Newman M 2018 {\em Networks\/} (Oxford University Press)

\bibitem{keeling2005networks}
Keeling M~J and Eames K~T 2005 {\em Journal of the Royal Society Interface\/}
  {\bf 2} 295--307

\bibitem{barthelemy2011spatial}
Barth{\'e}lemy M 2011 {\em Physics Reports\/} {\bf 499} 1--101

\bibitem{pastor2015epidemic}
Pastor-Satorras R, Castellano C, Van~Mieghem P and Vespignani A 2015 {\em
  Reviews of Modern Physics\/} {\bf 87} 925

\bibitem{porter2016dynamical}
Porter M~A and Gleeson J~P 2016 {\em Frontiers in Applied Dynamical Systems:
  Reviews and Tutorials\/} {\bf 4}

\bibitem{pastor2001epidemic}
Pastor-Satorras R and Vespignani A 2001 {\em Physical Review Letters\/} {\bf
  86} 3200

\bibitem{balcan2009multiscale}
Balcan D, Colizza V, Gon{\c{c}}alves B, Hu H, Ramasco J~J and Vespignani A 2009
  {\em Proceedings of the National Academy of Sciences\/} {\bf 106}
  21484--21489

\bibitem{brockmann2013hidden}
Brockmann D and Helbing D 2013 {\em Science\/} {\bf 342} 1337--1342

\bibitem{world2020coronavirus}
Organization W~H {\em et~al.\/} 2020

\bibitem{brandenburg2020quadratic}
Brandenburg A 2020 {\em arXiv preprint arXiv:2002.03638\/}

\bibitem{ziff2020fractal}
Ziff A~L and Ziff R~M 2020 {\em medRxiv preprint 2020.02.16.20023820\/}

\bibitem{li2020scaling}
Li M, Chen J and Deng Y 2020 {\em arXiv preprint arXiv:2002.09199\/}

\bibitem{maier2020effective}
Maier B~F and Brockmann D 2020 {\em Science\/} {\bf 368} 742--746

\bibitem{gross2020spatio}
Gross B, Zheng Z, Liu S, Chen X, Sela A, Li J, Li D and Havlin S 2020 {\em
  arXiv preprint arXiv:2003.08382\/}

\bibitem{manchein2020strong}
Manchein C, Brugnago E~L, da~Silva R~M, Mendes C~F~O and Beims M~W 2020 {\em
  arXiv preprint arXiv:2004.00044\/}

\bibitem{vazquez2006polynomial}
Vazquez A 2006 {\em Physical Review Letters\/} {\bf 96} 038702

\bibitem{ahn2007analysis}
Ahn Y~Y, Han S, Kwak H, Moon S and Jeong H 2007 Analysis of topological
  characteristics of huge online social networking services {\em Proceedings of
  the 16th International Conference on World Wide Web\/} pp 835--844

\bibitem{borgatti2018analyzing}
Borgatti S~P, Everett M~G and Johnson J~C 2018 {\em Analyzing social
  networks\/} (Sage)

\bibitem{medo2006distance}
Medo M 2006 {\em Physica A: Statistical Mechanics and its Applications\/} {\bf
  360} 617--628

\bibitem{riley2015five}
Riley S, Eames K, Isham V, Mollison D and Trapman P 2015 {\em Epidemics\/} {\bf
  10} 68--71

\bibitem{grassberger1983critical}
Grassberger P 1983 {\em Mathematical Biosciences\/} {\bf 63} 157--172

\bibitem{jia2020prediction}
Jia L, Li K, Jiang Y, Guo X {\em et~al.\/} 2020 {\em arXiv preprint
  arXiv:2003.05447\/}

\bibitem{wu2020generalized}
Wu K, Darcet D, Wang Q and Sornette D 2020 {\em arXiv preprint
  arXiv:2003.05681\/}

\bibitem{bodova2020emerging}
Bodova K and Kollar R 2020 {\em arXiv preprint arXiv:2005.06933\/}

\bibitem{chang2020modelling}
Chang S~L, Harding N, Zachreson C, Cliff O~M and Prokopenko M 2020 {\em arXiv
  preprint arXiv:2003.10218\/}

\bibitem{watts1998collective}
Watts D~J and Strogatz S~H 1998 {\em Nature\/} {\bf 393} 440

\bibitem{medo2008heterogeneous}
Medo M and Smrek J 2008 {\em The European Physical Journal B\/} {\bf 63}
  273--278

\bibitem{janssen2004generalized}
Janssen H~K, M{\"u}ller M and Stenull O 2004 {\em Physical Review E\/} {\bf 70}
  026114

\bibitem{lawson2016handbook}
Lawson A~B, Banerjee S, Haining R~P and Ugarte M~D 2016 {\em Handbook of
  spatial epidemiology\/} (CRC Press)

\bibitem{khaleque2013susceptible}
Khaleque A and Sen P 2013 {\em Journal of Physics A: Mathematical and
  Theoretical\/} {\bf 46} 095007

\bibitem{estrada2016epidemic}
Estrada E, Meloni S, Sheerin M and Moreno Y 2016 {\em Physical Review E\/} {\bf
  94} 052316

\bibitem{meyer2014power}
Meyer S and Held L 2014 {\em The Annals of Applied Statistics\/} {\bf 8}
  1612--1639

\bibitem{hohle2016infectious}
H{\"o}hle M 2016 Infectious disease modelling {\em Handbook of spatial
  epidemiology\/} (Chapman \& Hall/CRC)

\bibitem{daqing2011dimension}
Daqing L, Kosmidis K, Bunde A and Havlin S 2011 {\em Nature Physics\/} {\bf 7}
  481--484

\bibitem{bollobas2001random}
Bollob{\'a}s B and B{\'e}la B 2001 {\em Random graphs\/} (Cambridge University
  Press)

\bibitem{hethcote2000mathematics}
Hethcote H~W 2000 {\em SIAM Review\/} {\bf 42} 599--653

\bibitem{li2020substantial}
Li R, Pei S, Chen B, Song Y, Zhang T, Yang W and Shaman J 2020 {\em Science\/}

\bibitem{li2020early}
Li Q, Guan X, Wu P, Wang X, Zhou L, Tong Y, Ren R, Leung K~S, Lau E~H, Wong J~Y
  {\em et~al.\/} 2020 {\em New England Journal of Medicine\/}

\bibitem{bar2020sars}
Bar-On Y~M, Flamholz A~I, Phillips R and Milo R 2020 {\em arXiv preprint
  arXiv:2003.12886\/}

\bibitem{qian2020indoor}
Qian H, Miao T, Li L, Zheng X, Luo D and Li Y 2020 {\em medRxiv preprint
  2020.04.04.20053058\/}

\bibitem{singer2020short}
Singer H 2020 {\em arXiv preprint arXiv:2003.11997\/}

\bibitem{clauset2009power}
Clauset A, Shalizi C~R and Newman M~E 2009 {\em SIAM Review\/} {\bf 51}
  661--703

\end{thebibliography}
\end{document}